\documentclass{article}
\usepackage{lnfprep}
\usepackage{graphicx}
\usepackage{subfigure}
\usepackage{amsmath}
\usepackage{amssymb}

\def\beq{\begin{equation}}   \def\eeq{\end{equation}}
\def\bea{\begin{eqnarray}}   \def\eea{\end{eqnarray}}

\newcommand{\gsim}{\lower.7ex\hbox{$ \;\stackrel{\textstyle>}{\sim}\;$}}
\newcommand{\lsim}{\lower.7ex\hbox{$ \;\stackrel{\textstyle<}{\sim}\;$}}

\def\c2{CLEO~II.V}

\def\d0d0{ D^0\bar{D}^0 }
\def\p0p0{ P^0\bar{P}^0 }

\def\qp2{ \Bigl| \frac{q}{p} \Bigr|^2 }
\def\pq2{ \Bigl| \frac{p}{q} \Bigr|^2 }

\def\ps2s{  \psi(2S) }
\def\q2{ $q^2$ }
\def\cm2s1{ $\,{\rm cm}^{-2} {\rm s}^{-1}$}
\def\d0{D_2^{*0}}
\def\d+{D_2^{*+}}

\newcommand{\Header}{
  \begin{tabular}{rl}
  \hspace{-.4cm}
      &
    \renewcommand{\arraystretch}{0.5}
    \renewcommand{\arraystretch}{1}
  \end{tabular}
  \vskip 1cm
  \begin{flushright}
  \renewcommand{\arraystretch}{0.5}
    \begin{tabular}{r}
      {\underline{INFN-14-14/LNF}}\\    
      {\small 11 novembre 2014} \\      
    \end{tabular}
  \end{flushright}
  \renewcommand{\arraystretch}{1}
  \vskip 1 cm
  }
\begin{document}
\begin{titlepage}
\title{
  \Header{\LARGE \bf A study of HFO-1234ze (1,3,3,3-Tetrafluoropropene) as an eco-friendly replacement in RPC detectors}
 }
\author{
L. Benussi$^1$, S. Bianco$^1$, M. Ferrini$^2$, L. Passamonti$^1$, D. Pierluigi$^1$, D. Piccolo$^1$, A. Russo$^1$, G. Saviano$^2$   
} 

  \maketitle
  \baselineskip=1pt

\begin{abstract}
\indent 
The operations of Resistive Plate Chambers in LHC experiments require F-based
   gases for optimal performance. Recent regulations demand the use of
    environmentally unfriendly F-based gases to be limited or banned.
    This report shows results of studies on performance of RPCs operated with a potential eco-friendly  gas candidate  
1,3,3,3-Tetrafluoropropene, commercially known as HFO-1234ze.
\end{abstract}

\vspace*{\stretch{2}}

\vskip 1cm
\begin{flushleft}
\vskip 1cm
\begin{tabular}{l l}
  \hline\\
  $ ^{1\,\,\,}$& \footnotesize{Laboratori Nazionali di Frascati dell'INFN, Italy;} \\
  $ ^{2}$ & \footnotesize{Laboratori Nazionali di Frascati dell'INFN and Facolta' di Ingegneria Roma1, Italy} \\
 \\
\end{tabular}
\end{flushleft}
\end{titlepage}
\pagestyle{plain}
\setcounter{page}2
\baselineskip=17pt
\section{Introduction}  
\label{INTRO}
Resistive Plate Chambers (RPCs)~\cite{santonico}  are widley used in High Energy Phyiscs applications and in particular in the LHC experiments. 
For applications where high background rates are expected they have to be operated in avalanche mode in order to keep the total produced charge
low with benefits in terms of aging and rate capability. This is usually obtained with suitable gas mixtures that prevents the transition from
avalanche to stremer keeping the detection efficiency above 90 \%. The use of F-based gases, usually used in refrigerants, have shown 
so far to give the best performance.
The RPC system of the CMS~\cite{cms} experiment is operated with a gas mixture of $C_{2}H_{2}F_{4}$  95.2 \%, isobuthane 4.5 \% and $SF_{6}$ 0.3 \% showing 
very high performance in the LHC environment~\cite{rpcperformance}.
Recent european regulations demand the use of environmentally un-friendly F-based gases to be limited or banned. 
The impact of a refrigerant on the environment has to be quantified in terms of the contribution to the greenhouse effect and to the 
depletion of the ozone layer. The first mentioned effect is measured in Global Warmth Potential (GWP), and is normalized to the effect of $CO_{2}$  (GWP = 1),
 and the effect on the ozone layer is measured in Ozone Depletion Potential (ODP), normalized to the effect of $CCl_{3}F$ (ODP = 1). 
The European Community has prohibited the production and use of gas mixtures with Global Warming Power $>$ 150. 

The $C_{2}H_{2}F_{4}$  and $SF_{6}$ gases used in the RPCs for example present a GWP=1430 and 23900 respectively 
and need to be replaced with components with lower GWP.

An overview of potential gas candidates can be found in~\cite{jogvan}. 
In this note we will concentrate on the use of HFO1234ze component in the RPC gas mixture. First results of RPCs operated with this component and in 
streamer mode have been already presented recently~\cite{liberti}. 
In this note independent tests are performed and additional gas combinations are investigated. 

 \section{Experimental setup} 
  \label{SETUP}

\begin{figure}[Ht] 
	\centering
	\includegraphics[width=0.7\textwidth]{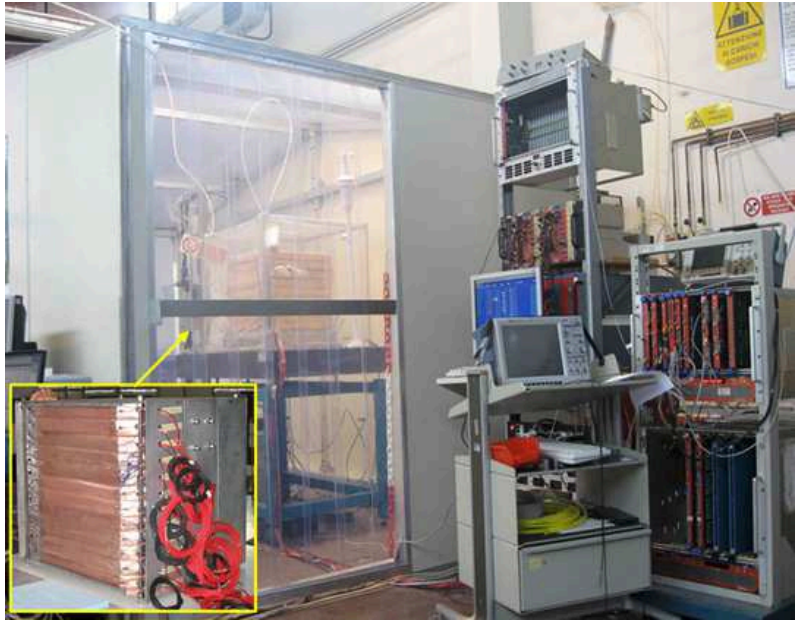}
	\caption{Schematic layout of the GGM system of the CMS RPC detector reproduced in the Frascati Laboratories.}
	\label{fig:SETUP}
	\end{figure}

 An array composed of 12 single-gap square (50 x 50) $cm^2$ RPC detectors is in operation at Frascati national laboratories of INFN
and arranged as a cosmic ray hodoscope. The system is a copy of the Gas Gain Monitoring (GGM) system of the Compact Muon Solenoid (CMS)
 RPC muon detector at the Large Hadron Collider (LHC) of CERN, Geneva, Switzerland described elsewhere~\cite{Benussi:2008fp}~\cite{Colafranceschi:2012qf}.
The system is located inside a T, RH controlled box and flushed with different gas mixtures  humidified at about 40\% (Fig.\ref{fig:SETUP}). 

\begin{figure}[Htbp]
	\centering
	\includegraphics[width=0.4\textwidth]{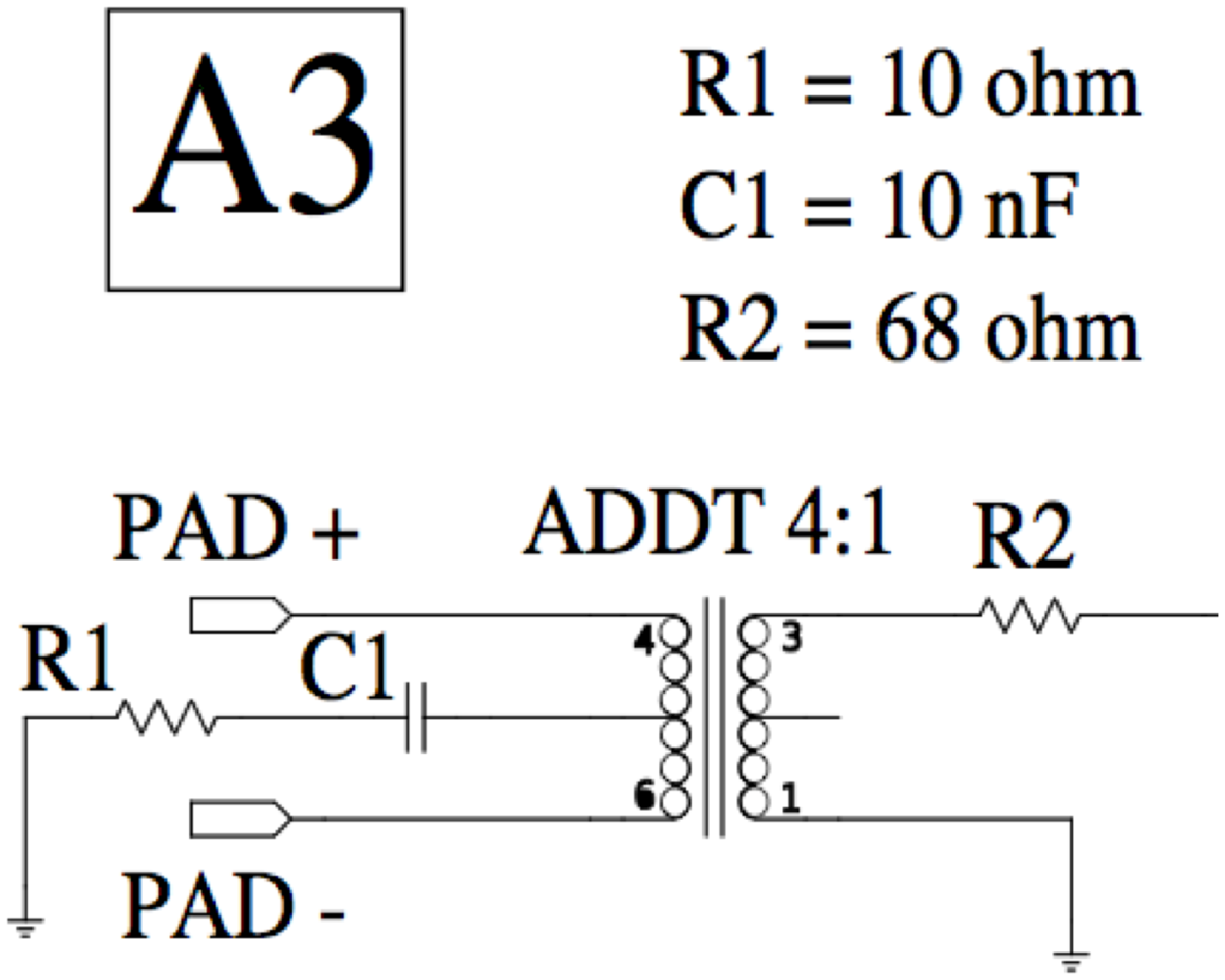}
	\caption{Scheme of double pad readout of RPCs used for the tests}
	\label{fig:doublePad}
	\end{figure}

Each chamber of the GGM system consists of a single gap with double sided pad read- out: two copper pads are glued on 
the two opposite external side of the gap. Two foam planes are used to reduce the capacity coupling between the pad and the copper 
shields used as screen for the system. The signal is read-out by a transformer based circuit (fig.~\ref{fig:doublePad}) that allows to algebraically 
subtract the two signals, which have opposite polarities, and to obtain an output signal with subtraction of the coherent noise, 
with an improvement of about a factor 4 of the signal to noise ratio.
	 The trigger is provided by scintillator layers located on the top and bottom
of the RPC stack with an average event rate of about 1 Hz, corresponding to about 20 minutes for 1000
	 events.

\section{DAQ and signal readout}
\label{DAQ}

Measurements of the RPC performance as a function of different gas mixtures are obtained with a simple readout of the signals with a digital oscilloscope. 
In the tests performed so far we acquired the signals from only two RPCs. Both detectors are operated with the same gas mixture but one of the two
is maintained at a fixed working voltage and used as additional offline trigger to the events. 

The signal pick-up extracted from the pad of each detector is sent to an oscilloscope lecroy with 5 GSamples and 1 GHz band width externally triggered by 
the scintillator system. Measurements are taken at different Effective High Voltages normalized at the pressure $P_{0}$ = 990 mbar and temperature 
$T_{0}$= 290 $^\circ K$ according to the standard formula~\ref{povert}.

\begin{equation}
HV_{eff} = HV \times P_{0}/P \times T/T_{0}
\label{povert}
\end{equation}
 For each HV setting about 1000 waveforms are saved on the disk for offline analysis.

 In order to reduce the environmental noise picked up from the system a 20 Mhz filter is applied on each RPC channel signal. 
The use of the filter in addition to the fact that we read a pad electrode, modifies the signal shape increasing the raising and falling time of the 
electric pulse. The overall induced charge is anyway not affected by this shape modifications and the timing of the signals is weakly affected by using 
a low electronic threshold to define the time of the signal.

\begin{figure}[Htbp]
\centering
\includegraphics[width=0.7\textwidth]{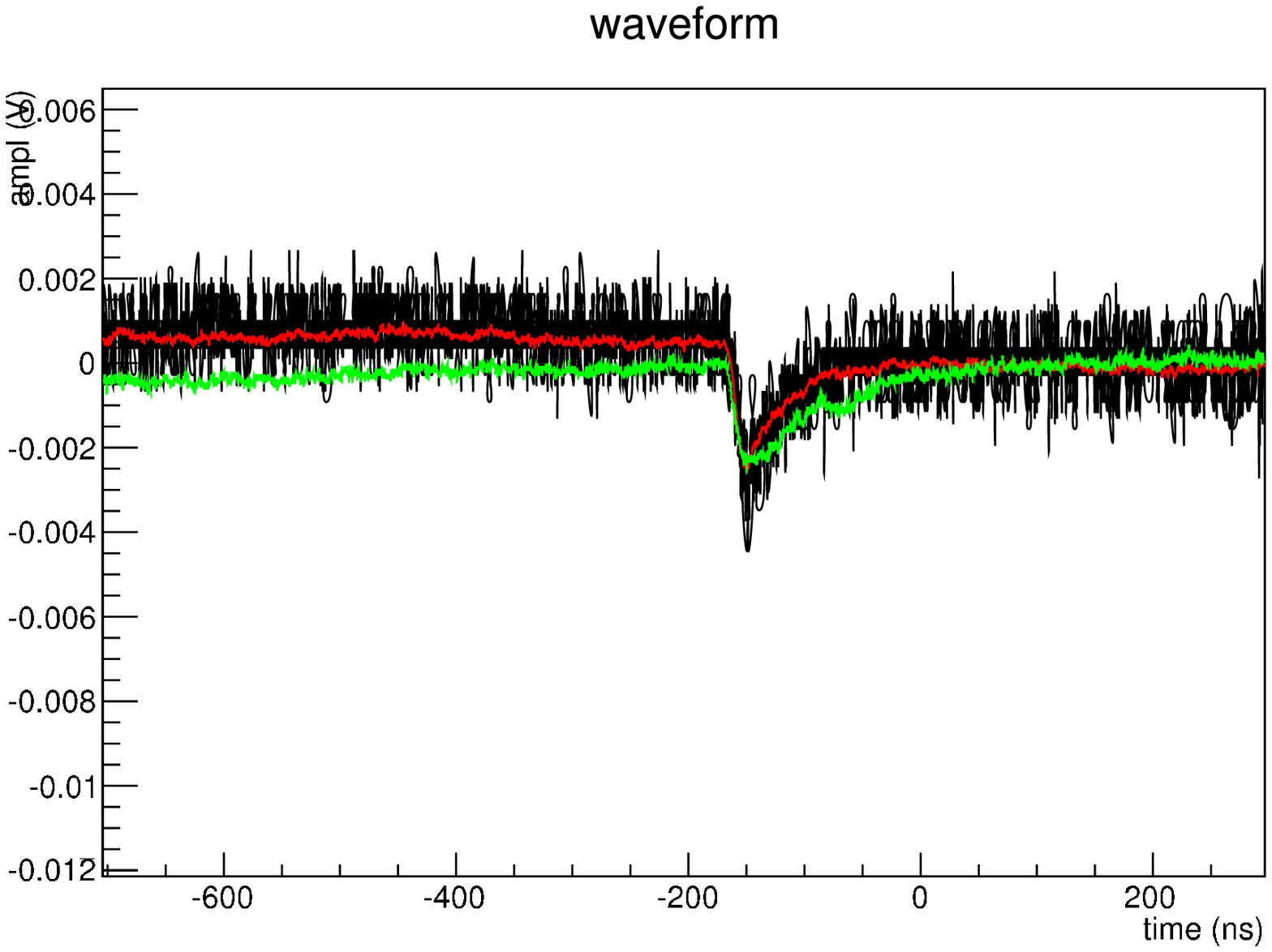}
\caption{Typical RPC signals stored from oscilloscope. One RPC signal is registred in two diferent channels with different y scale (red and black tracks).
The green track corresponds to the trigger RPC signal.}
\label{fig:pedagogicalExample}
\end{figure}

 In figure \ref{fig:pedagogicalExample} a typical set of waveforms is shown. 
A total time window of about 1 $\mu$s is acquired and for analysis purpose it is divided in four regions.

The Region of 150 ns around the trigger is the "signal window" and is used to measure the signal characteristics. The 10 ns region before the 
"signal window" ("baseline window") is used for baseline evaluation. For each of the three channels the voltage baseline is evaluated averaging 
the voltage measurements in this region 
in which signals are absent and then subtracted to the measurement in the "signal window". The 150 ns   before the "baseline window" defined as 
"control region" is used to check the width of the charge distribution and of the minimum voltage in absence of signals in order to define a correct 
threshold to identify RPC signals. Finally the 150 ns above the "signal region" (``delayed region'') is used to identify delayed  
streamers.

The data are saved on the disk in form of ascii files storing the time-voltage waveform history. Analysis is performed offline with ROOT software.

\begin{figure}[Ht] 
	\centering
	\includegraphics[width=0.45\textwidth]{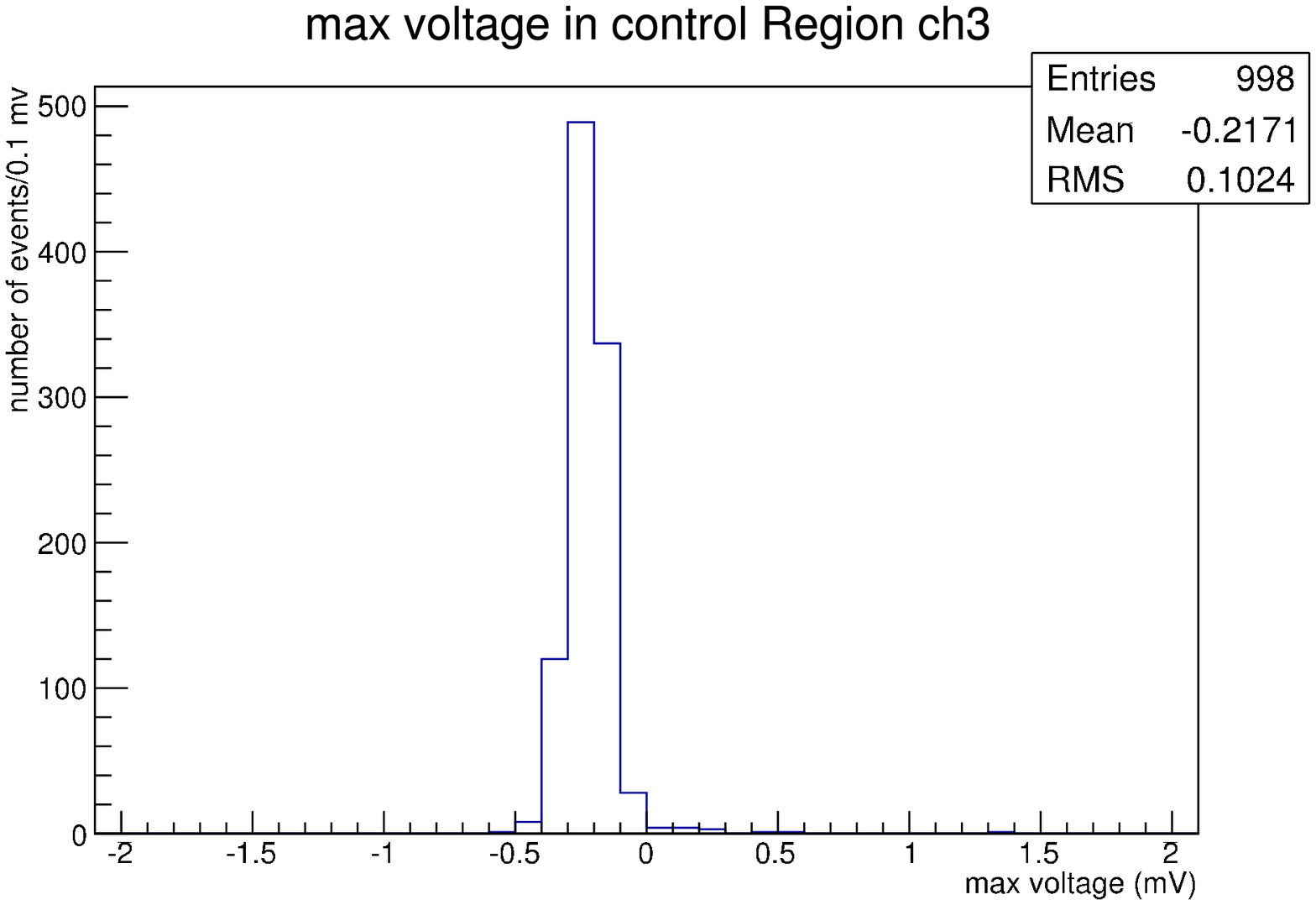}
	\includegraphics[width=0.45\textwidth]{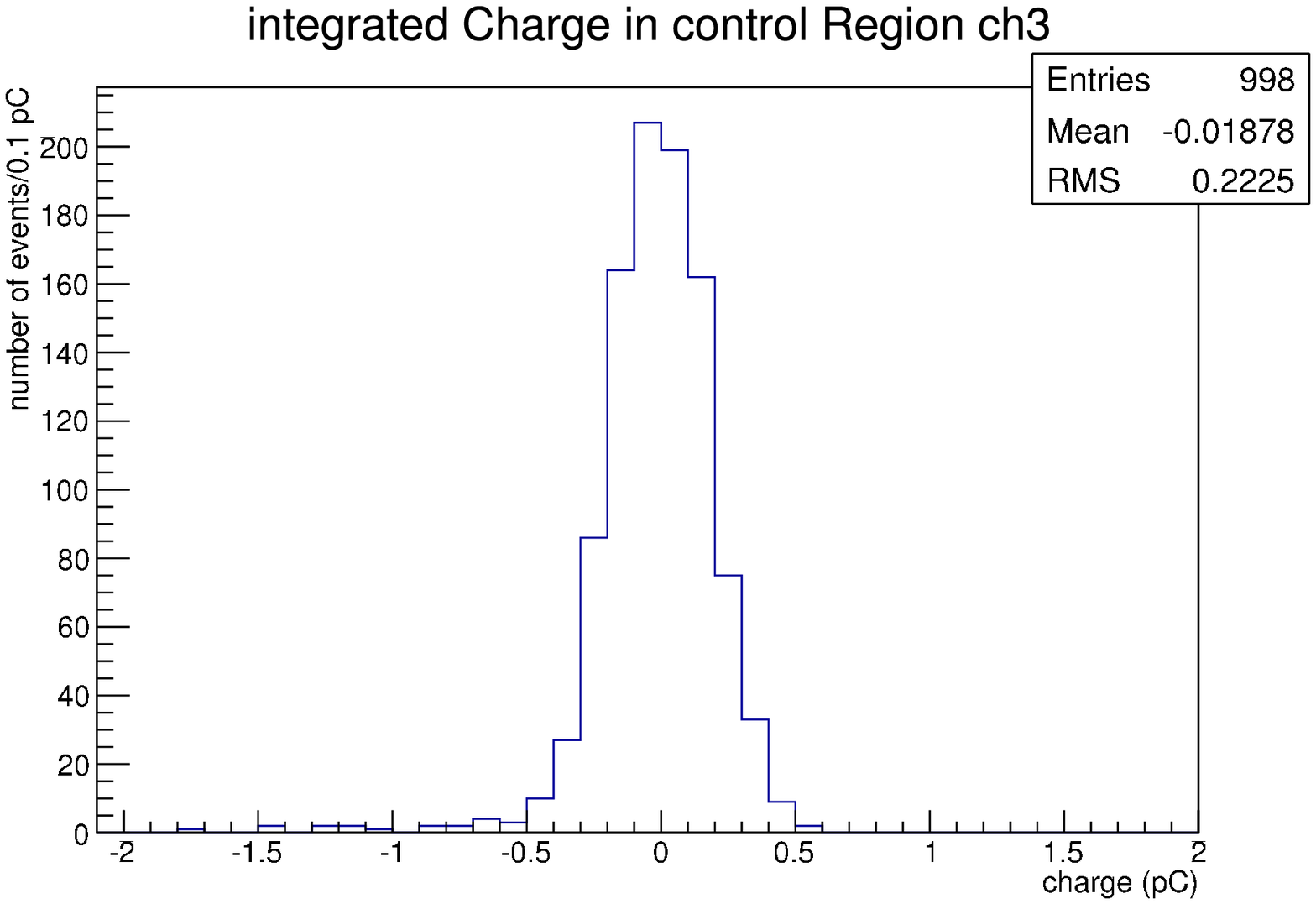}
	\caption{Left: Distribution of the minimum of the voltage in the ``control region''. A threshold of -0.4 mV is enough to discriminate fake
signals due to noise induced baseline variations. Right: Distribution of the induced charge integrated in a 150 ns time window of the ``control region''. 
A threshold of 0.3 pC in addition to the minimum voltage threshold reduce the fake signals and simulate an electronic threshold similar 
to that used in the CMS experiment. }
	\label{fig:maxVoltageControlRegion}
	\end{figure}

\begin{figure}[Htbp]
	\centering
	\includegraphics[width=0.45\textwidth]{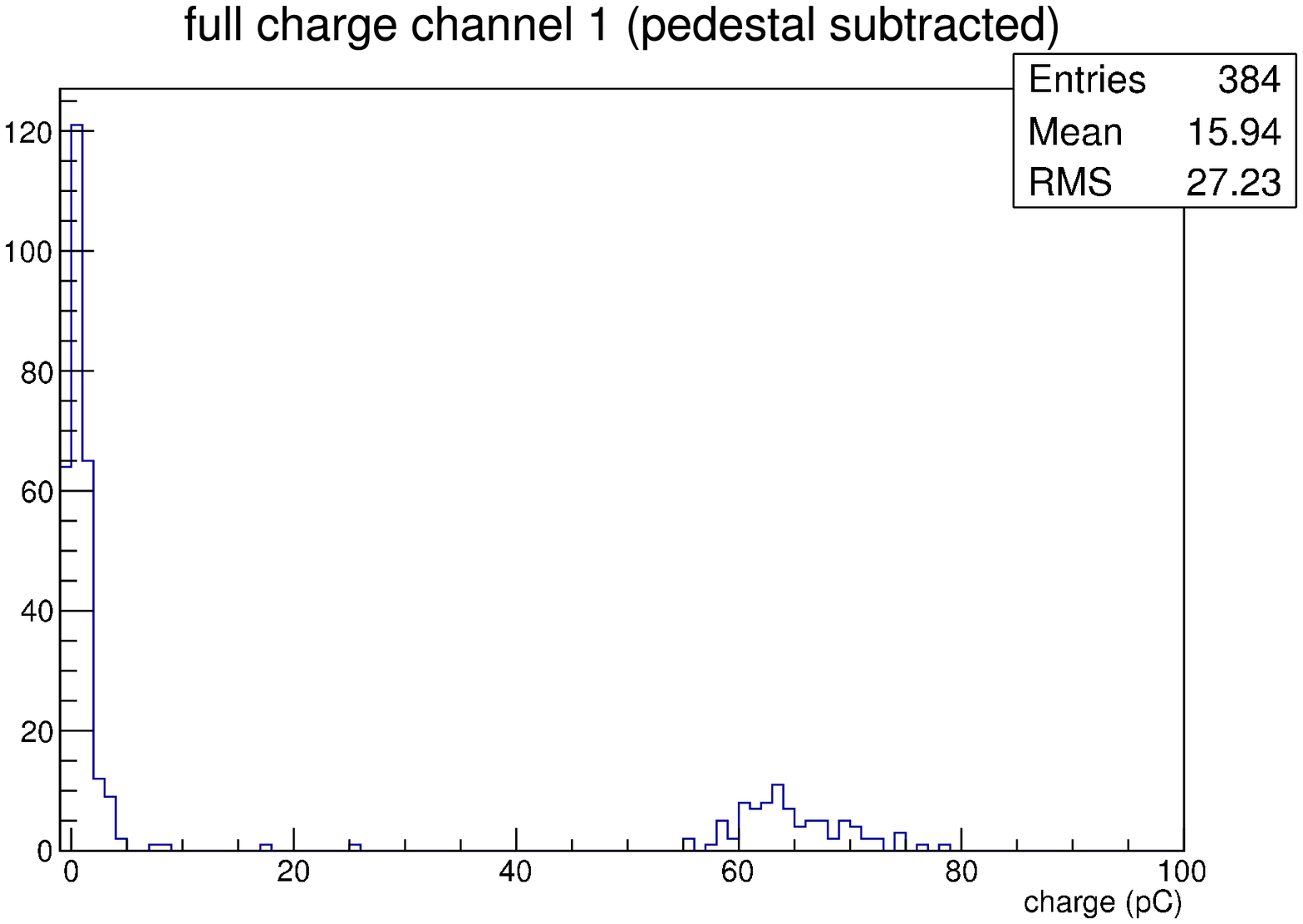}
	\includegraphics[width=0.45\textwidth]{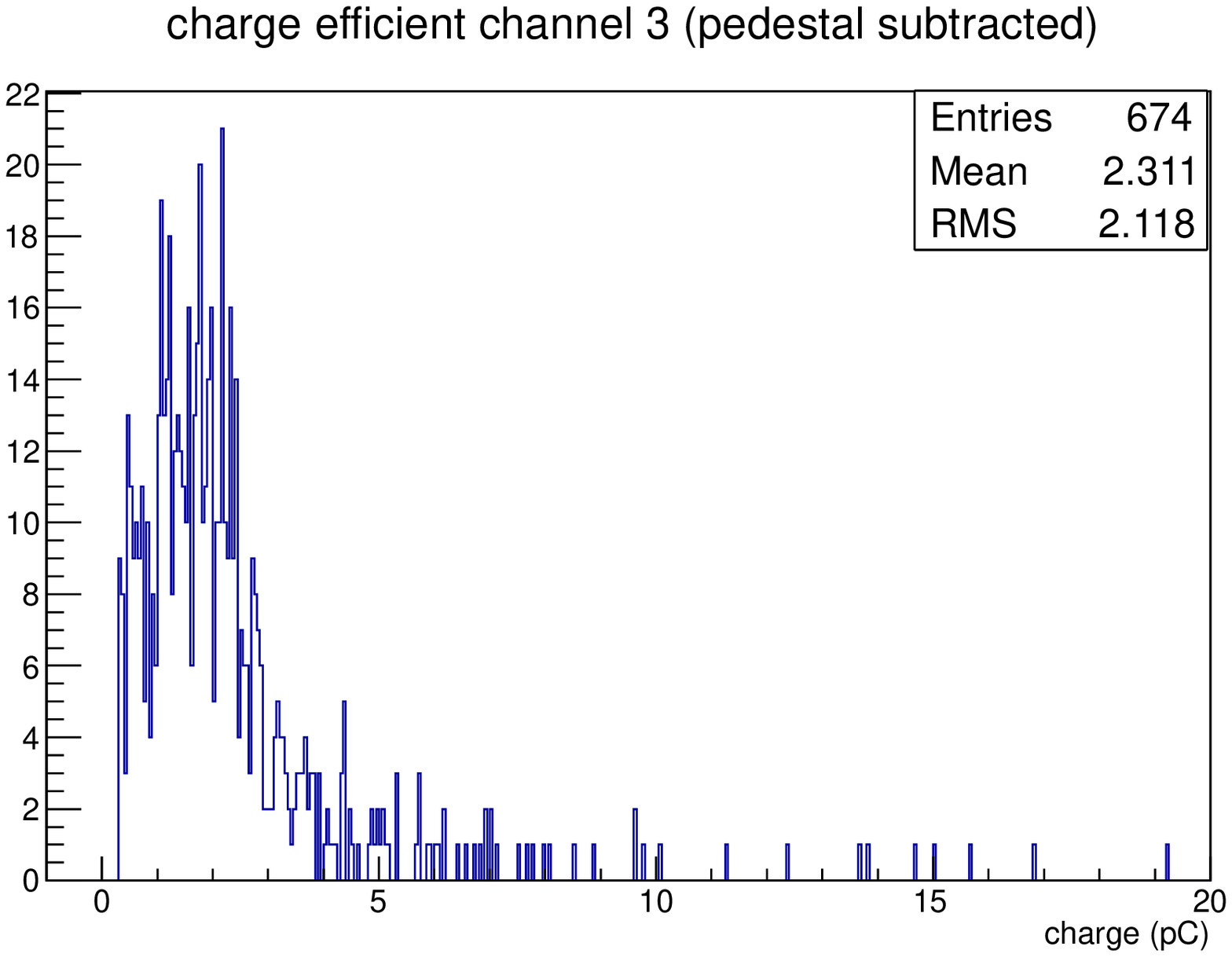}
	\caption{Left: Typical distribution of the induced charge integrated in a 150 ns time window of the ``signal region'' for one of the gas mixtures
tested and working in streamer mode. The peak of the streamers is clearly seen for induced charges above 20 pC. Right: induced charge distribution 
for the CMS standard gas mixture showing the avalanche peak}
	\label{fig:totalChargeSignalRegion}
	\end{figure}

For each event after having corrected the measurements in the "signal window" for  the baseline extracted from the "baseline window"
 the integral of the signal in the 150 ns "signal window" is converted in pC in order to plot the induced charge. 
In figures \ref{fig:maxVoltageControlRegion} (left) and \ref{fig:maxVoltageControlRegion} (right)
it is shown  the minimum  voltage and the integral of the charge, after the baseline correction, in the  "control region" while in 
figure \ref{fig:totalChargeSignalRegion} (left) and  \ref{fig:totalChargeSignalRegion} (right) the induced charge in the "signal region" respectively 
for gas mxiture with high streamer probability and for the CMS standard gas mixture. The two different horizontal scales are used to
better identify the avalanche and streamer peaks.
As can be seen from the plot of fig.\ref{fig:maxVoltageControlRegion} in the control region a threshold of -0.4 mV on the minimum voltage plot is enough 
to exclude fake signals due to pedestal variations and to take the RPC signals. In addition we also require that the 
integrated charge is above 0.3 pC that is very close to the present threshold of the electronics used in the CMS experiment~\cite{feb}.\footnote{The threshold
applied in CMS ranges between 150 and 175 fC, but in our layout the induced charge is about twice the charge induced on CMS strips becauses of the double
 pad readout.}

To evaluate the streamer probability we define a streamer as  a signal with integrated charge above 20 pC in the 300 ns time window defined by the 
"signal" + "delayed" regions. 
Finally the time resolution is obtained measuring the time over a threshold of -0.4 mV of signals coming from two different RPCs and plotting the difference. 
The values 
coming  from the distribution should then be divided for a factor of the order of $\sqrt{2}$ in order to  extract the time resolution. 

Gas mixtures are prepared via gas flowmeters and then analyzed in output to the RPC system with a gas chromatograph to evaluate the tested  working mixture.
In all the tests we started from the standard gas mixture used in the RPC CMS system ($C_{2}H_{2}F_{4}$  95.2 $\%$, isobuthane 4.5 $\%$, $SF_{6}$ 0.3 $\%$) and then we 
change the amount of $C_{2}H_{2}F_{4}$ , $SF_{6}$ and add fractions of HFO1234ze and Argon keeping almost unchanged the isobuthane content. However small variations between the expected and the
measured fraction of each component can be found. For each test we quote the mixture measured with gas chromatograph.

\section{results on $SF_{6}$ scan}
\label{sfsix}

\begin{figure}[Htbp]
	\centering
	\includegraphics[width=0.8\textwidth]{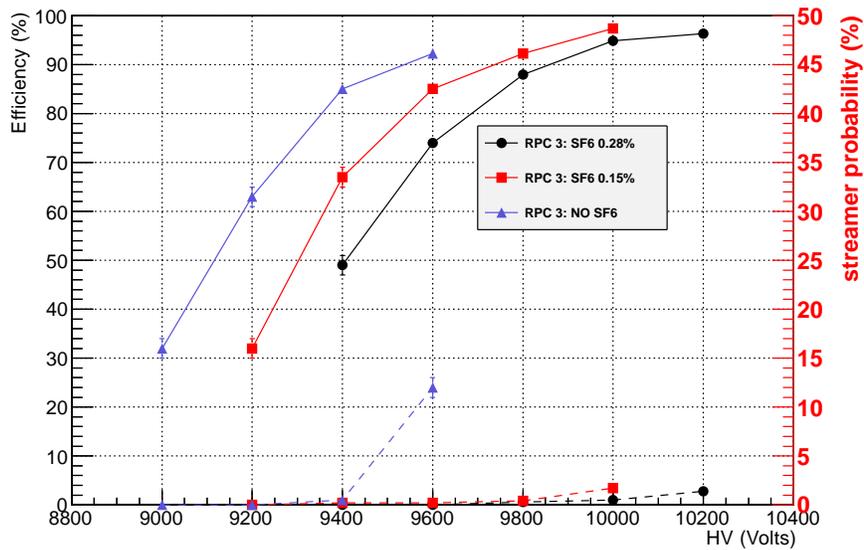}
	\caption{Efficiency and streamer probability vs effective HV for a mixture composed by 4.5 \% of isobuthane, about 95.2 \% of R134a and varying 
the amount of $SF_{6}$ between 0 and 0.28 \%. The amount of R134a is adjust accordingly. 
Streamer probability is indicate with dashed line and the scale is on the right y axis}
	\label{fig:EfficiencyComparison_SF6}
	\end{figure}

\begin{figure}[Htbp]
	\centering
	\includegraphics[width=0.5\textwidth, angle=-90]{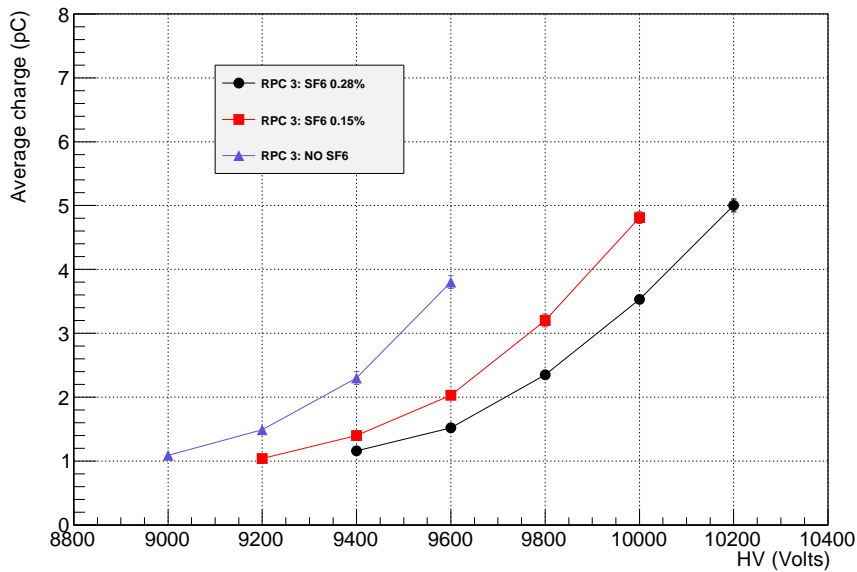}
	\caption{Integrated induced charge vs effective HV for a misture composed by 4.5 \% of isobuthane, about 95.2 \% of R134a and varying the amount 
of $SF_{6}$ between 0 and 0.28 \%. The amount of R134a is adjust accordingly}
	\label{fig:ChargeComparison_SF6}
	\end{figure}

\begin{figure}[Htbp]
	\centering
	\includegraphics[width=0.8\textwidth]{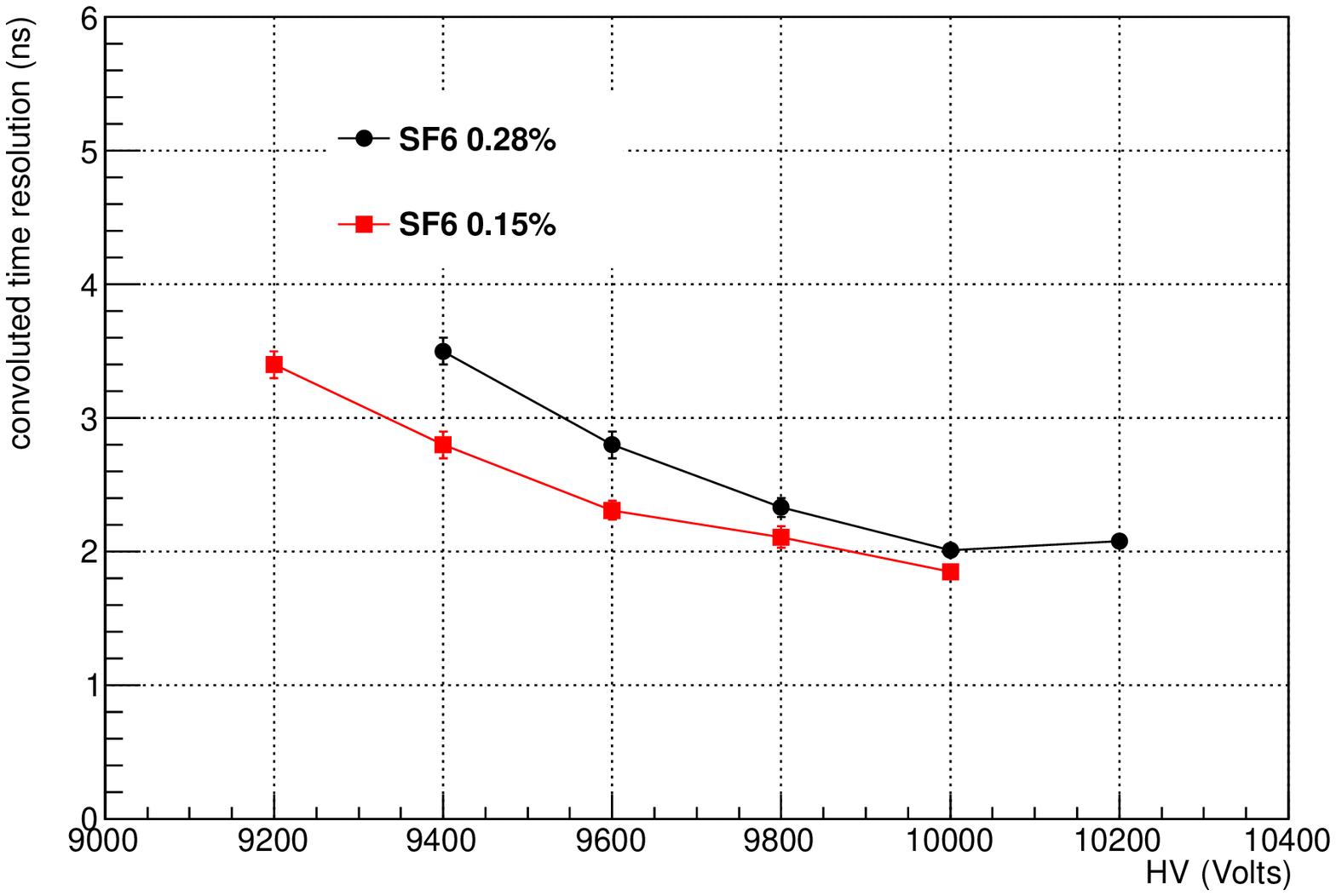}
	\caption{Difference between the time over threshold of two RPCs vs effective HV for a misture composed by 4.5 \% of isobuthane, about 95.2 \% of R134a and varying the amount of $SF_{6}$ between 0.15 and 0.28 \%. The amount of R134a is adjust accordingly. One of the two RPCs is fixed at 9.2 kV while
the HV of the RPC under test is changed. To get the real time resolution the values should be divided for something of the order of $\sqrt{2}$. }
	\label{fig:timeResComparison_SF6}
	\end{figure}

In order to validate the method a first set of measurement have been performed starting from the CMS standard gas mixture and varying the 
amount of $SF_{6}$. 
 Figures \ref{fig:EfficiencyComparison_SF6}, \ref{fig:ChargeComparison_SF6} and \ref{fig:timeResComparison_SF6} show
respectively  the efficiency and streamer probability, the induced charge and the time over threshold difference between the RPC under test and a reference RPC at fixed voltage 
as a function of High voltage for the different gas mixtures tested.
As expected reducing the amount of $SF_{6}$ in the mixture the working point is shifted to lower values but the streamer probability increase.

\section{Adding HFO-1234ze to the gas mixture}
\label{hfo}

\begin{figure}[Htbp]
	\centering
	\includegraphics[width=0.5\textwidth, angle=-90]{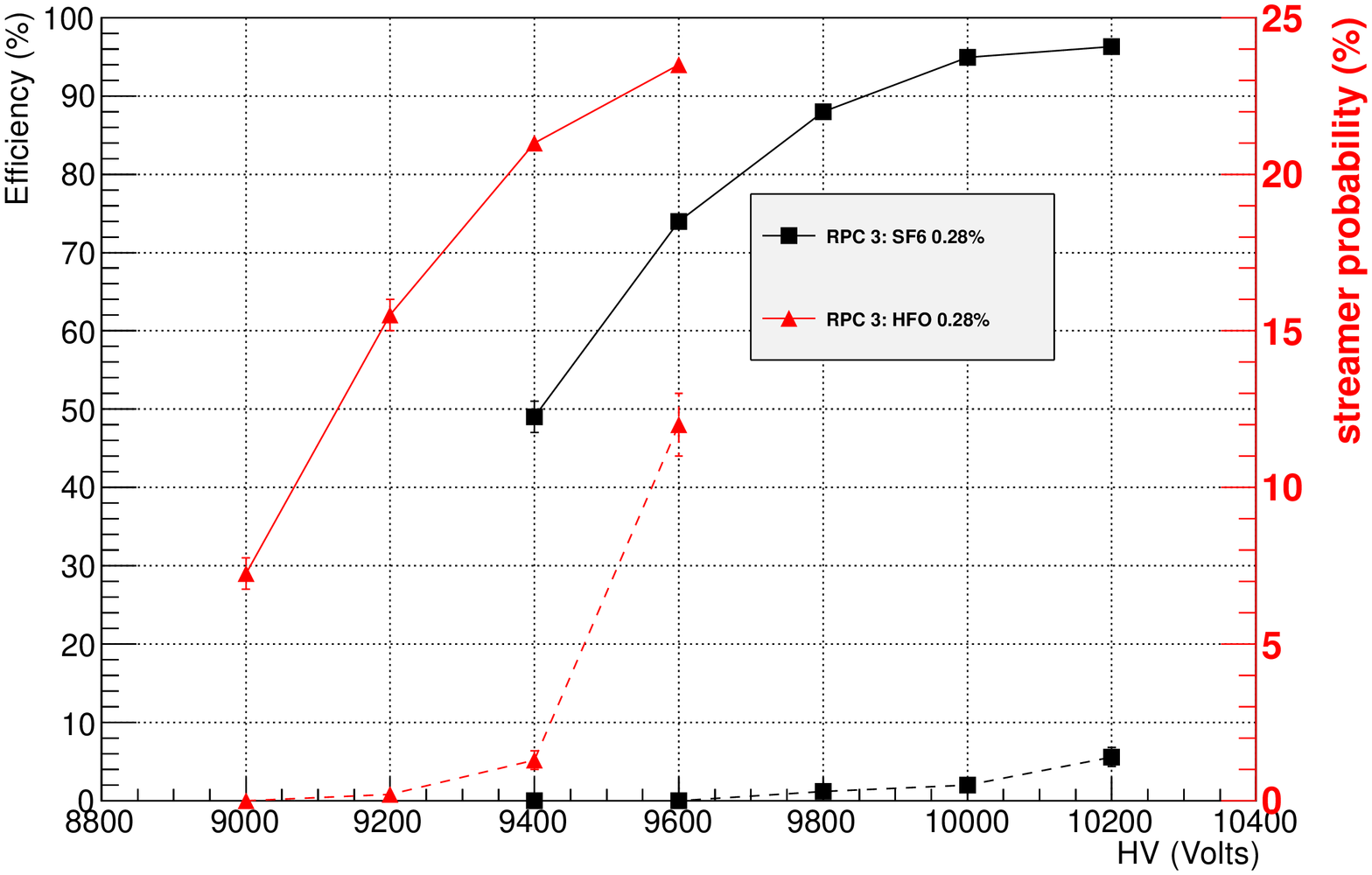}
	\caption{Efficiency and streamer probability vs effective HV for a mixture composed by 4.5 \% of isobuthane, 95.2 \% of R134a 
and $SF_{6}$ 0.28 \% in one case anf HFO1234ze 0.28\% in a second case. Results are shown for two chambers.}  
	\label{fig:EfficiencyComparison_SF6vsHFO}
	\end{figure}

\begin{figure}[Htbp]
	\centering
	\includegraphics[width=0.5\textwidth, angle=-90]{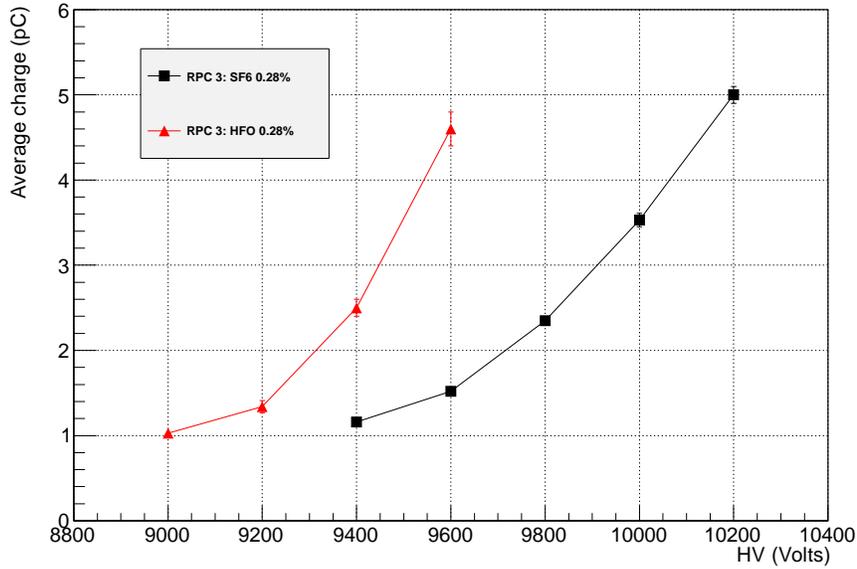}
	\caption{Integrated induced charge vs effective HV for a misture composed by 4.5 \% of isobuthane, 95.2 \% of R134a
and $SF_{6}$ 0.28 \% in one case anf HFO1234ze 0.28\% in a second case. Results are shown for two chambers.}
	\label{fig:ChargeComparison_SF6vsHFO}
	\end{figure}

\begin{figure}[Htbp]
	\centering
	\includegraphics[width=0.5\textwidth, angle=-90]{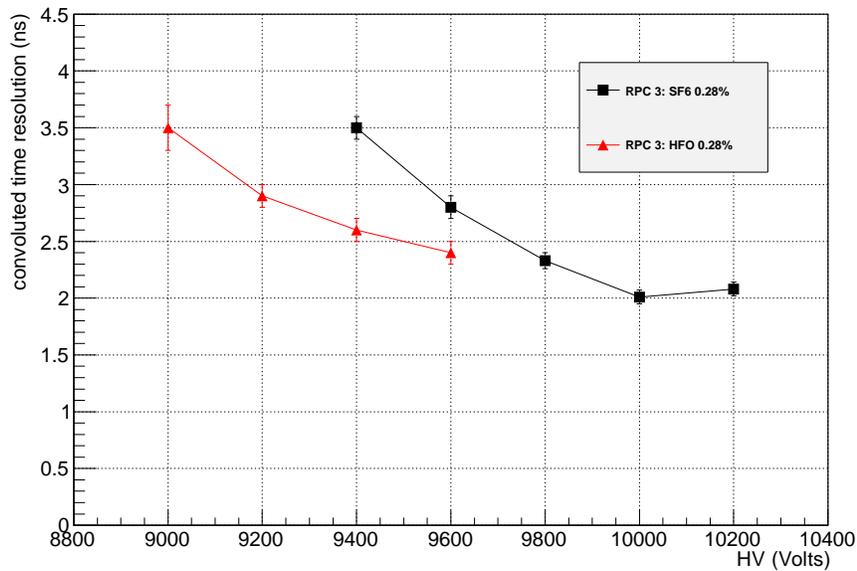}
	\caption{Difference between the time over threshold of two RPCs vs effective HV for a misture composed by 4.5 \% of isobuthane, 95.2 \% of R134a 
and $SF_{6}$ 0.28 \% in one case anf HFO1234ze 0.28\% in a second case. One  RPC is fixed at 9.2 kV while
the HV of other two RPCs under test is changed. To get the real time resolution the values should be divided for something of the order of $\sqrt{2}$. }
	\label{fig:timeResComparison_SF6vsHFO}
	\end{figure}

 To evaluate the impact of the new HFO1234ze component in the RPC gas mixture in a first test we replaced the $SF_{6}$ with the HFO. Efficiency/streamer probability, induced charge and time resolution of two RPCs  are compared respectively 
in fig  \ref{fig:EfficiencyComparison_SF6vsHFO} \ref{fig:ChargeComparison_SF6vsHFO} \ref{fig:timeResComparison_SF6vsHFO} for standard CMS 
gas mixture and for the case in which $SF_{6}$ has been replaced by the HFO. 
The addition of a small quantity of HFO does not have the same impact as the $SF_{6}$ for what concerns streamer suppression. 

\begin{figure}[Htbp]
	\centering
	\includegraphics[width=0.5\textwidth, angle=-90]{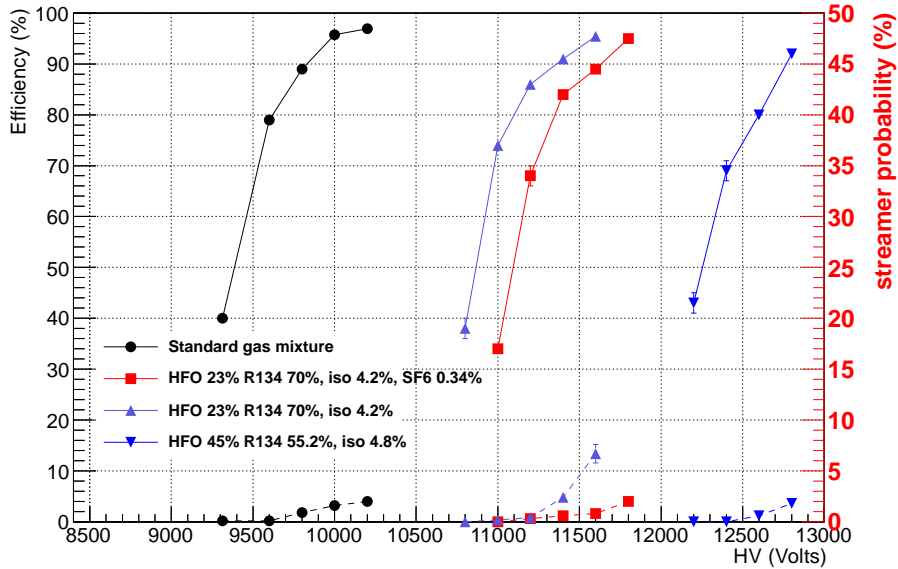}
	\caption{Efficiency and streamer probability vs effective HV for different mistures composed by isobuthane, R134a, HFO1234ze with and without $SF_{6}$. Streamer probability is indicate with dashed line and the scale is on the right y axis}
	\label{fig:EfficiencyComparison_R134_HFO}
	\end{figure}

\begin{figure}[Htbp]
	\centering
	\includegraphics[width=0.5\textwidth, angle=-90]{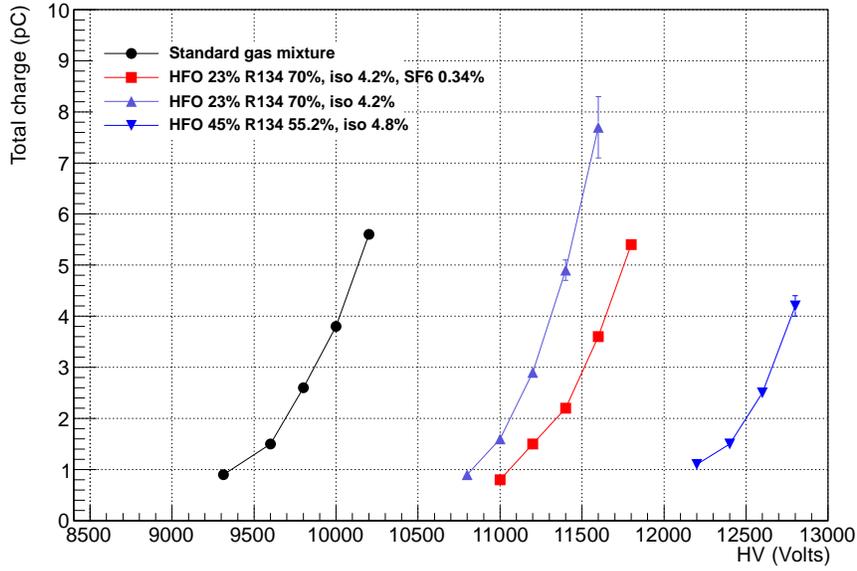}
	\caption{Integrated induced charge vs effective HV for different mistures composed by isobuthane, R134a, HFO1234ze with and without $SF_{6}$ }
	\label{fig:ChargeComparison_R134_HFO}
	\end{figure}

\begin{figure}[Htbp]
	\centering
	\includegraphics[width=0.5\textwidth, angle=-90]{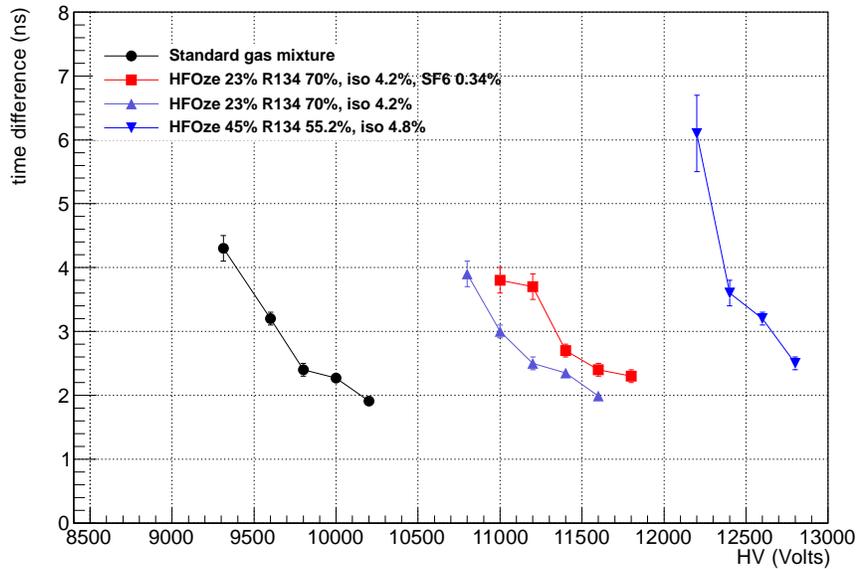}
	\caption{Difference between the time over threshold of two RPCs vs effective HV  for different mistures composed by isobuthane, R134a, HFO1234ze with and without $SF_{6}$. To get the real time resolution the values should be divided for something of the order of $\sqrt{2}$. }
	\label{fig:timeResComparison_R134_HFO}
	\end{figure}

A second set of measurement have been performed with gas mixtures in which isobuthane has been maintained fixed around 
4.5 $\%$ and varying the amount of HFO-1234ze vs $C_{2}H_{2}F_{4}$ and the $SF_{6}$ content. Figures \ref{fig:EfficiencyComparison_R134_HFO},
 \ref{fig:ChargeComparison_R134_HFO} and \ref{fig:timeResComparison_R134_HFO}, show the impact of the addition of 23$\%$ and 45$\%$ of
HFO-1234ze replacing $C_{2}H_{2}F_{4}$ , with and without $SF_{6}$ addition. An amount of about 23$\%$ is enough to move the working point of about 1500 Volts
while the fraction of streamers is maintained low if $SF_{6}$ is added to the mixture. 

 In any case the HFO1234ze component shows improved quenching characteristics with respect to R134a and 
replacing the full amount of $C_{2}H_{2}F_{4}$  with HFO-1234ze would require to increase the working voltage too much stressing the HV system 
so that our tests stopped at a fraction of 45$\%$ HFO-1234ze.

\section{Adding Argon to the HFO-1234ze based gas mixture}
\label{ArHfo}

We explored the effect of adding Argon to the HFO-1234ze based gas mixture in order to reduce the working point and eliminate the $C_{2}H_{2}F_{4}$  component
in the mixture. Previews results~/cite{ref:liberti} shown that a mixture of Argon with small percentage of HFO-1234ze give encouraging results in streamer 
mode. We started from the already tested mixtures and then increased the fraction of HFO-1234ze to analyze the performance.

\begin{figure}[t]
          \centering
	  \includegraphics[width=0.5\textwidth, angle=-90]{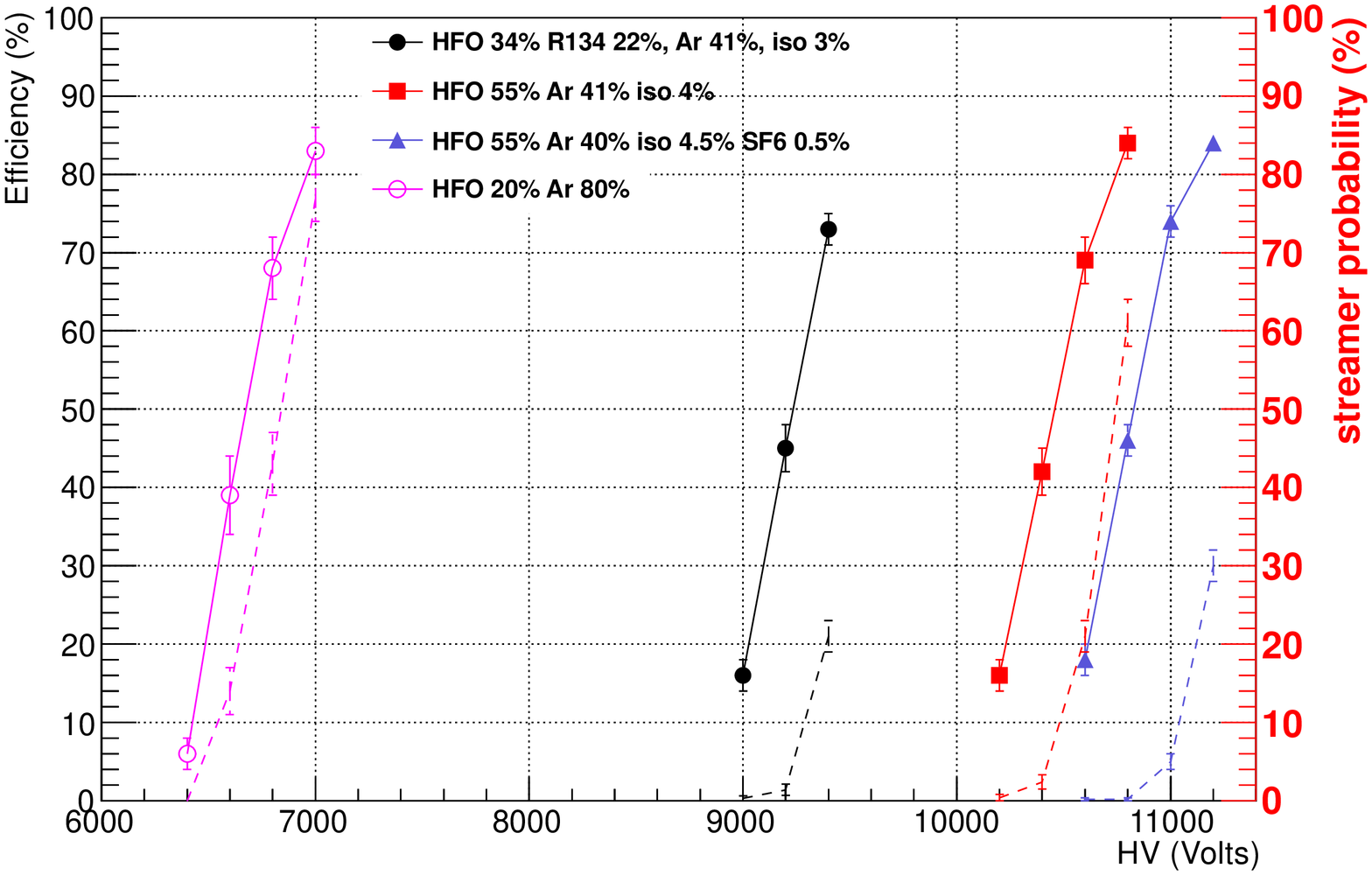}
	  \caption{Efficiency and streamer probability vs effective HV for different mixtures adding Argon to the HFO1234ze based mixtures. Streamer probability is indicate with dashed line and the scale is on the right y axis}
	  \label{fig:EfficiencyComparison_Ar_HFO}
	\end{figure}

\begin{figure}[t]
          \centering
	  \includegraphics[width=0.5\textwidth, angle=-90]{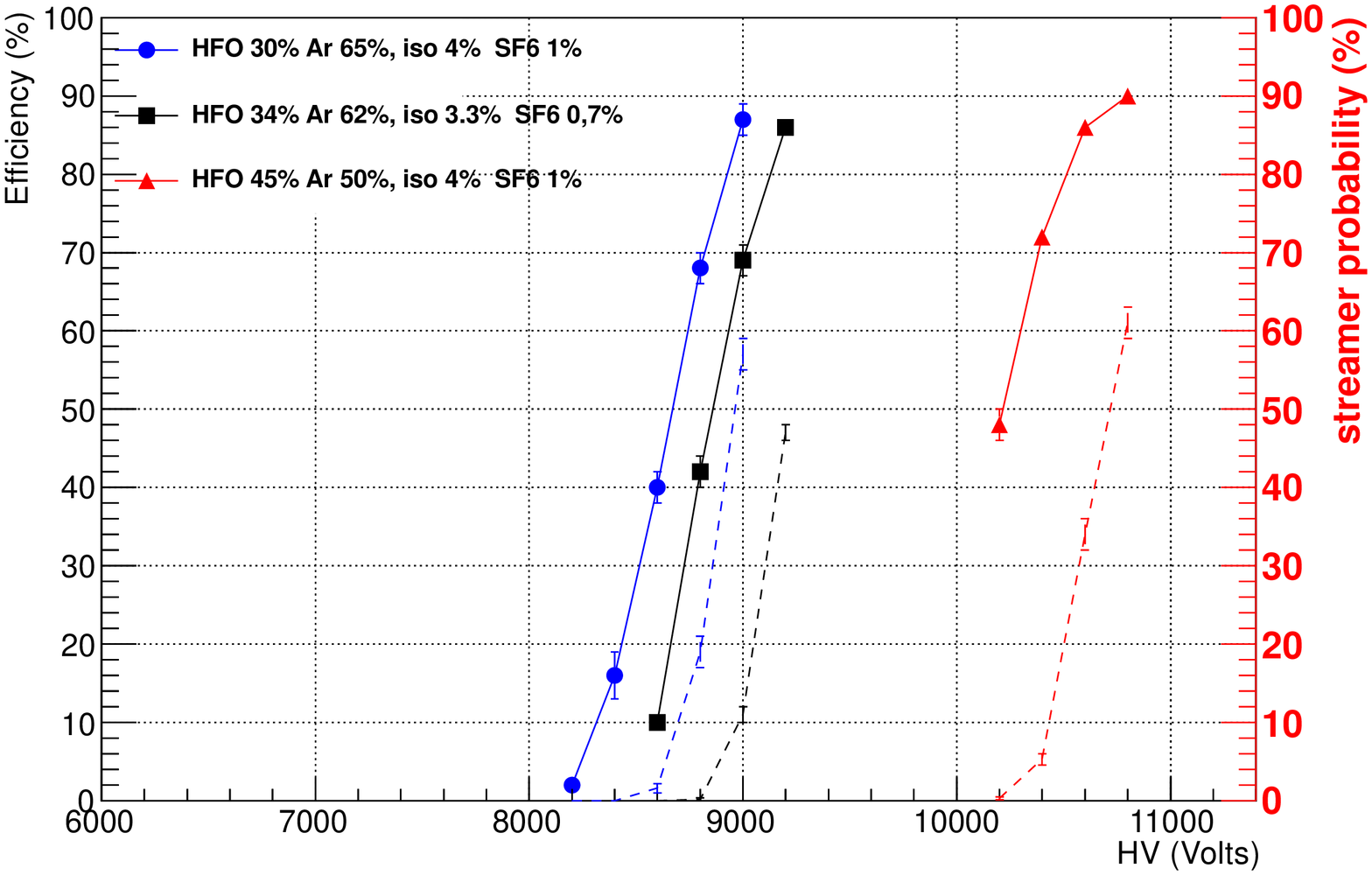}
	  \caption{Efficiency and streamer probability vs effective HV for different mixtures adding Argon to the HFO1234ze based mixtures. Streamer probability is indicate with dashed line and the scale is on the right y axis}
	  \label{fig:EfficiencyComparison_Ar_HFO_2}
	\end{figure}

\begin{figure}[t]
          \centering
	  \includegraphics[width=0.5\textwidth, angle=-90]{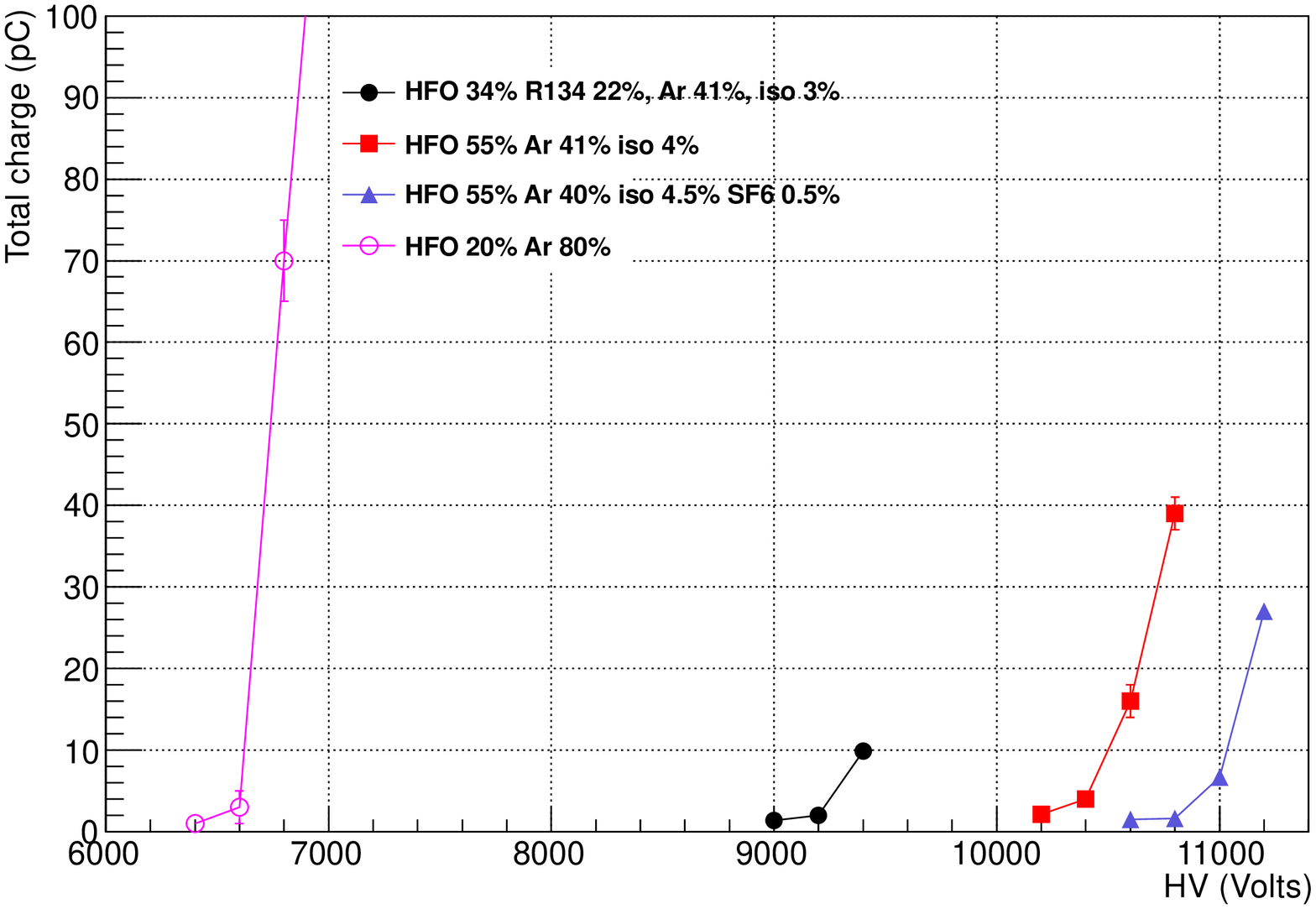}
	  \caption{Intergrated induced charge vs effective HV for different mixtures adding Argon to the HFO1234ze based mixtures. Streamer probability is indicate with dashed line and the scale is on the right y axis}
	  \label{fig:ChargeComparison_Ar_HFO}
	\end{figure}

\begin{figure}[t]
          \centering
	  \includegraphics[width=0.5\textwidth, angle=-90]{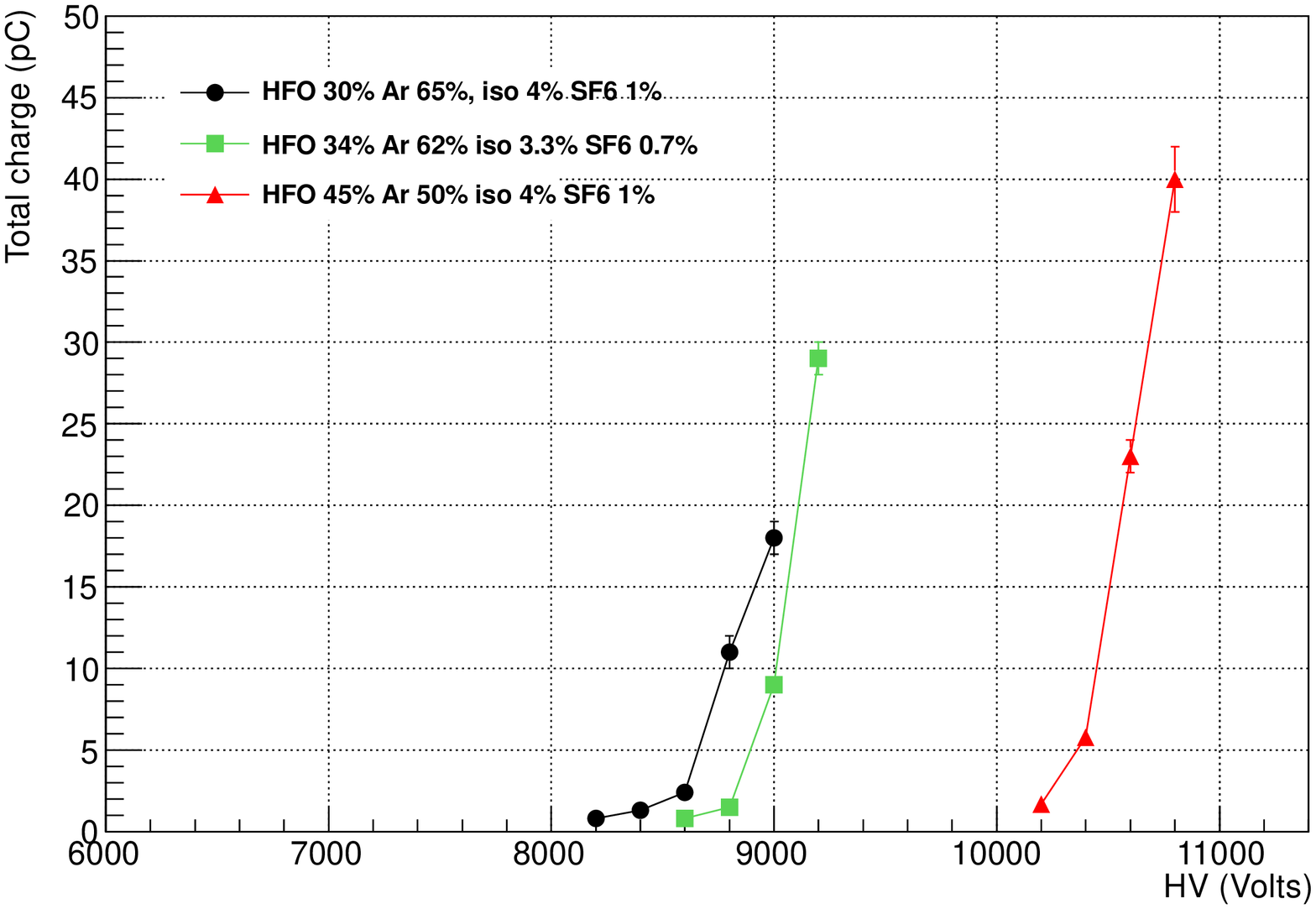}
	  \caption{Intergrated induced charge vs effective HV for different mixtures adding Argon to the HFO1234ze based mixtures. Streamer probability is indicate with dashed line and the scale is on the right y axis}
	  \label{fig:ChargeComparison_Ar_HFO_2}
	\end{figure}

\begin{figure}[t]
          \centering
	  \includegraphics[width=0.5\textwidth, angle=-90]{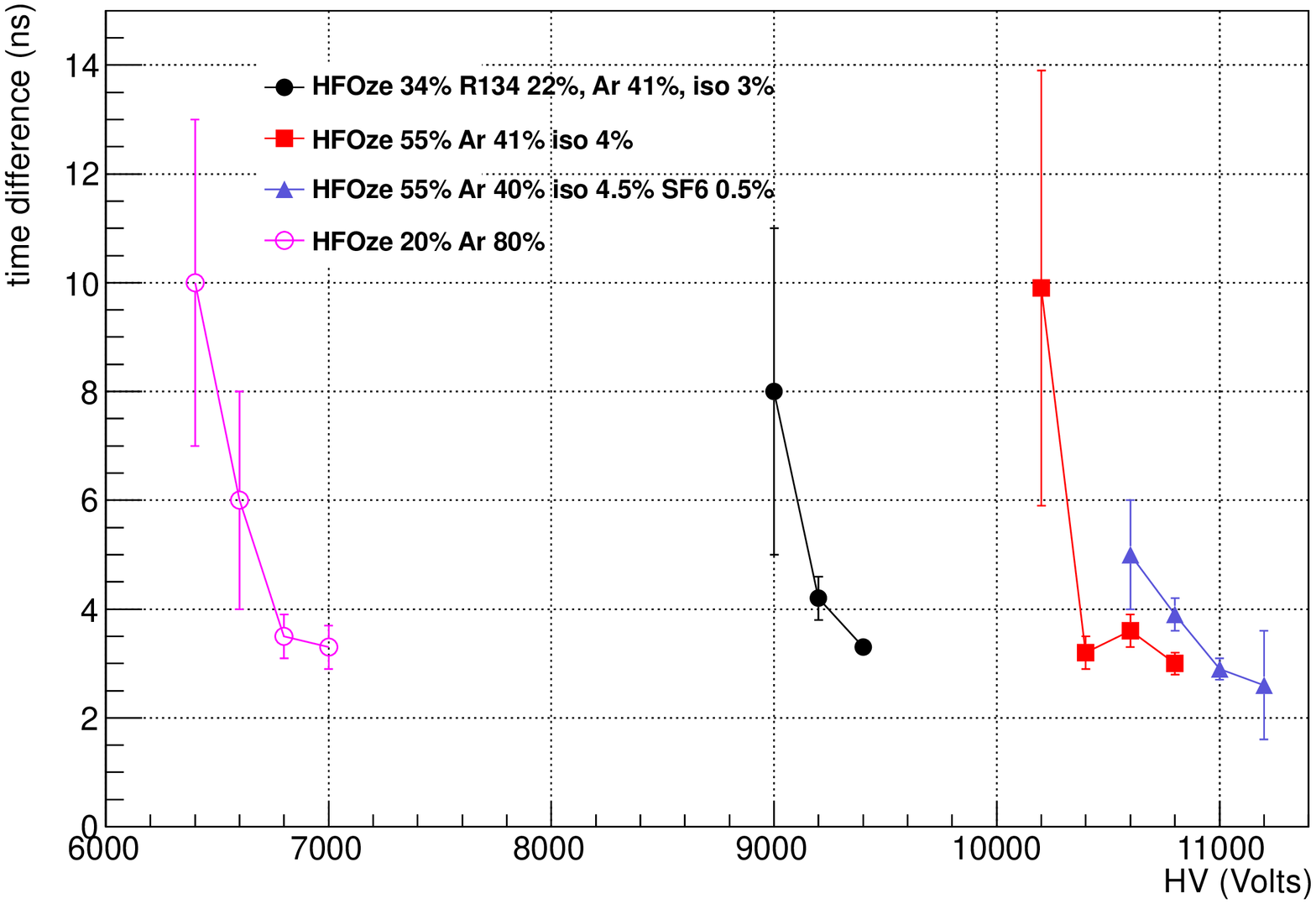}
	  \caption{Difference between the time over threshold of two RPCs vs effective HV   for different mixtures adding Argon to the HFO1234ze based mixtures. Streamer probability is indicate with dashed line and the scale is on the right y axis}
	  \label{fig:timeResComparison_Ar_HFO}
	\end{figure}

\begin{figure}[t]
          \centering
	  \includegraphics[width=0.9\textwidth]{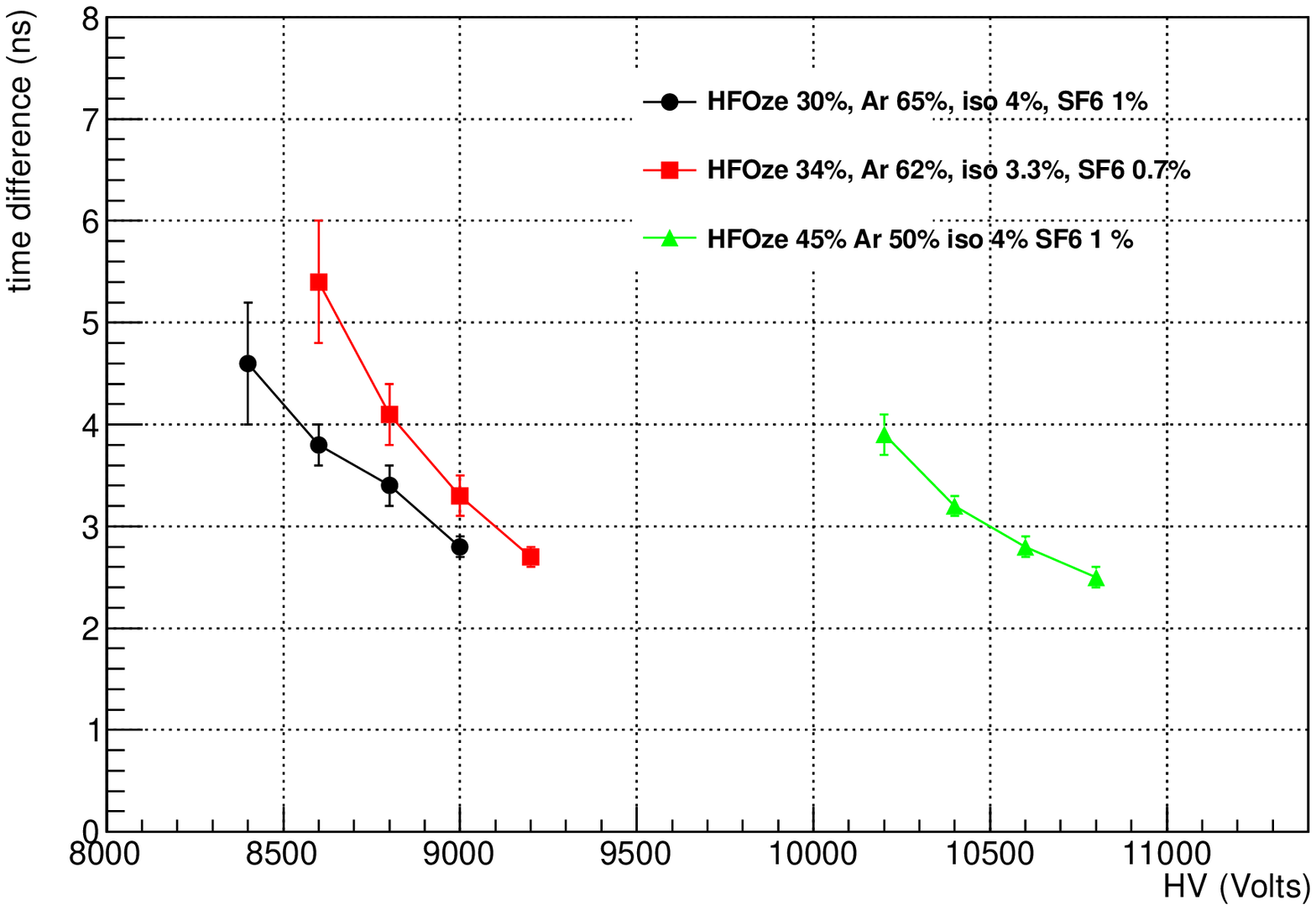}
	  \caption{Difference between the time over threshold of two RPCs vs effective HV   for different mixtures adding Argon to the HFO1234ze based mixtures. Streamer probability is indicate with dashed line and the scale is on the right y axis}
	  \label{fig:timeResComparison_Ar_HFO_2}
	\end{figure}

Figures \ref{fig:EfficiencyComparison_Ar_HFO},  \ref{fig:EfficiencyComparison_Ar_HFO_2}, \ref{fig:ChargeComparison_Ar_HFO}, \ref{fig:ChargeComparison_Ar_HFO_2}, \ref{fig:timeResComparison_Ar_HFO} and \ref{fig:timeResComparison_Ar_HFO_2} show as usual 
efficiency/streamer probability, induced charge and time resolution for different mixture compositions. Addition of HFO-1234ze
move as usual the working point to higher values but is not sufficient to reduce the fraction of streamers keeping the efficiency above 90$\%$ at 
least with  the tested electronic threshold. Reducing the electronic threshold could recover detection efficiency in the range where streamer probability is still low, but such tests were not possible with the present setup due to the environmental noise limiting the threshold to aplly.




\section{Conclusions}
\label{CONCLUSIONS}

Performance of 2 mm wide single gap Resistive Plate Chambers, operated with several HFO1234ze based gas mixtures, have been presented here in terms of 
efficiency, streamer probability, integrated charge and time resolution. The HFO1234ze shows large quenching characteristics that could be interesting
for RPC operations. At the same time it cannot be used to replace the R134a in large fractions as the working point is shifted to very high voltages.
Several mixtures with different fractions of HFO1234ze and Argon have been also tested giving incouraging results but still the full efficiency is note reached in absence of large streamer fractions, at least with the electronic thresholds used in CMS. More tests will be necessary to fully
characterize this new gas component for his use inside RPCs.

\newpage
%

%
\end{document}